 \documentclass[preprint]{aastex}

\shorttitle{Population gradients in Local Group dSphs}
\shortauthors{Harbeck et al.}

\begin{document}

\title{Population Gradients in Local Group Dwarf Spheroidals}

\author{ Daniel Harbeck\altaffilmark{1}, 
         Eva K. Grebel\altaffilmark{1,11},
	 Jon Holtzman\altaffilmark{2}, 
         Puragra Guhathakurta\altaffilmark{3}, 
         Wolfgang Brandner\altaffilmark{4,5}, 
         Doug Geisler\altaffilmark{6,11}\,, 
         Ata Sarajedini\altaffilmark{7,11}, 
         Andrew Dolphin\altaffilmark{8}, 
         Denise Hurley-Keller\altaffilmark{9,10,11}, 
         Mario Mateo\altaffilmark{10,11} }

\altaffiltext{1}{Max-Planck-Institut f\"ur Astronomie, K\"onigstuhl 17, 
                 D-69117 Heidelberg, Germany}
\altaffiltext{2}{New Mexico State University, Department 4500, Box 30001,
                 Las Cruces, NM 88003}
\altaffiltext{3}{UCO/Lick Observatory, 
                University of California at Santa Cruz, Santa Cruz, 
                CA 95064, USA}
\altaffiltext{4}{University of Hawaii, Institute for Astronomy, 
                 2680 Woodlawn Dr., Honolulu, HI 96822}
\altaffiltext{5}{European Southern Observatory, Karl-Schwarzschild-Str. 2, 
                 D-85748 Garching, Germany}
\altaffiltext{6}{Universidad de Concepci\'on, Grupo de Astronomia,
                 Departamento de F\'\i sica, Casilla 160-C,
                 Concepci\'on, Chile}
\altaffiltext{7}{University of Florida, Department of Astronomy, 
                 211 Bryant Space Sciences Center Gainesville, FL
                 32611}
\altaffiltext{8}{NOAO, 950 N. Cherry Avenue, Tucson, AZ 85726}
\altaffiltext{9}{Case Western Reserve University, 10900 Euclid Ave.,
	         Cleveland, OH 44106}
\altaffiltext{10}{University of Michigan, Department of Astronomy,
		 821 Dennison Building, Ann Arbor, MI 48109-1020}

\altaffiltext{11}{Visiting Astronomer, Cerro Tololo Inter-American
                  Observatory, National Optical Astronomy
                  Observatories, operated by the Association of
                  Universities for Research in Astronomy, Inc., under
                  cooperative agreement with the National Science
                  Foundation.}

\begin{abstract}
We present a systematic and homogeneous analysis of population
gradients for the Local Group dwarf spheroidals (dSphs) Carina,
Sculptor, Sextans, Tucana, Andromeda I-III, V, and VI. For all of the
Milky Way companions studied here we find significant population
gradients. The same is true for the remote dSph Tucana located at the
outskirts of the LG.  Among the M\,31 dSph companions only Andromeda I
and VI show obvious gradients. In all cases where a HB morphology
gradient is visible, the red HB stars are more centrally
concentrated. The occurence of a HB morphological gradient shows a
correlation with a morphology gradient in the red giant branch. It
seems likely that metallicity is the driver of the gradients in
Sextans, Sculptor, Tucana, and Andromeda VI, while age is an important
factor in Carina. We find no evidence that the vicinity of a nearby
massive spiral galaxy influences the formation of the population
gradients.
\end{abstract}


\section{Motivation}

Dwarf spheroidal (dSph) galaxies contribute to the faint end of the
galaxy luminosity function and may be basic building blocks in
hierarchical galaxy formation scenarios. In all Local Group (LG) dSphs
an old stellar population (age$\ge 10$~Gyr) exists as indicated by the
presence of a horizontal branch (HB).  The HBs of the dSphs mostly
show a bimodal morphology, i.e., both the blue and the red part of the
HB are populated. Over the last few years it has become clear that
these apparently simple galaxies actually show a large range of
metallicities as well as star to star variations in elemental
abundance ratios, indicating that dSphs have expirienced a complicated
star formation history with various self enrichment processes
\citep{mighell96,hurley98,gallart99,dolphin01, shetrone01}.

For a few LG dSphs radial gradients in the HB morphology have been
reported (e.g., Da Costa et al. 1996; Hurley-Keller, Mateo, \& Grebel
1999). In each case the blue HB (BHB) stars have a more extended
distribution than the red HB (RHB) stars.

Population gradients are well known for some dwarf irregular (dIrr)
galaxies with ongoing star formation, e.g., Sextans A, WLM, Leo A, and
IC 1613 as summarized by \citet{grebel99iau}. While high-mass dwarf
galaxies, e.g. the LMC, tend to show widespread, multiple zones of
concurrent star formation, low-mass dwarf galaxies ($10^7$~M$_\odot$)
show centrally concentrated younger populations \citep{grebel00}.  In
particular the transition type dIrr/dSph galaxy Phoenix and the dSph
Fornax show a central concentration of the younger stars
\citep{martinez99, stetson98}.

Radial gradients in the HB morphology can be interpreted as an imprint
of the ancient formation process of the dSphs and may help to
distinguish different formation scenarios. In this paper we perform a
homogeneous study of HB morphology gradients for a large sample of
dSphs in the Local Group. We analyze the spatial distribution of BHB
and RHB stars, the red giant branch (RGB) stars, and, if present, of
red clump (RC) stars of LG dSphs.

\section{The horizontal branch as a tracer of population gradients}

A HB forms in stellar populations older than $\sim 10$~Gyr and
consists of He core burning stars. BHB stars lie blueward of the RR
Lyrae strip, while the RHB stars have lower surface temperatures and
lie redwards of the instability strip.  The first parameter governing
the HB morphology of a population is metallicity: Metal rich
populations tend to form a red HB, while metal poor populations in
general form blue HBs. However, globular clusters with the same
metallicity may exhibit very different HB morphologies, demonstrating
that additional parameters play a role. For instance, differences in
age on a time scale of a few Gyr can have a significant influence on
the HB morphology \citep{sarajedini97}. For a fixed metallicity the HB
morphology can change from a completely red HB to an entirely blue HB
for population ages between $11$~Gyr and $15$~Gyr
\citep{lee94,lee01}. Populations with ages between $1$ and $10$~Gyr
form an RC instead of a HB \citep{girardi98}. A red clump is indeed
the continuation of the sequence of core He-burning stars, but is
formed by stars with higher total masses. The existence of an RC
therefore is a tracer of intermediate age stellar populations in a
galaxy. Consequently, the existence of spatial variations in the
distribution of core He-burning stars (RHB, BHB, and RC) can reflect
either age or metallicity differences, or a mixture of both. Other
parameters may still play a r\^ole. For example, the old, metal rich
open cluster NGC 6791 contains a population of stars on the extended
blue HB in addition to its expected red HB
\citep{kaluzny92,liebert94}.

In an attempt to separate the effects of age and metallicity, we also
will consider variations in the RGB morphology. For a given age, a
lower metallicity results in a bluer RGB. A decrease in age has a
similar effect (age-metallicity degeneracy). We demonstrate this
effect in Fig.~\ref{fig_iso} where we show isochrones
\citep{girardi00} for $z=0.004$ and $z=0.0004$ for ages of  $10$, and
$14$~Gyr. Note that for ages larger than $6$~Gyr the effect of age on
the RGB morphology is nearly negligible.

\section{Data sample and reduction}

The LG contains 17 currently known dSphs. We consider here 3 of the 9
Milky Way companions, namely Carina, Sculptor, and Sextans, and 5 of
the 6 Andromeda dSphs, Andromeda I-III, V, and VI. In addition we
analyze the isolated dSph Tucana. Our sample is restricted to those
galaxies for which deep, wide-field data are available to us. To study
the distribution of the blue and red HB stars, observations of a
sufficiently large field and deep photometry are required. We present
results from our own observations as well as data from the Hubble
Space Telescope (HST) archive. The data sources of our sample are
summarized in Tab.~\ref{tab_obslog}, including the instrument used,
filter set, and radius covered, and references where available. We
show in Fig.~\ref{fig_chart} the observed field of view for each dSph
on a Digital Sky Survey image.

All HST observations (the five Andromeda dSphs and Tucana; see
Tab.~\ref{tab_obslog}) were reduced using HSTphot
\citep{dolphin00}. This program performs point spread function
photometry on Wide Field Planetary Camera 2 (WFPC\,2)
\citep{trauger94} observations with PSFs based loosely on Tiny Tim
models \citep{krist93}. It carries out corrections for charge transfer
efficiency effects and performs the photometric calibration and
transformation to the Johnson photometric system
\citep{dolphin00b}. The photometry of the five Andromeda dSphs and
Tucana were dereddened according to the extinction maps of
\citet{schlegel}. The Andromeda dSphs were observed with the F450W and
F550W filters. \citet{dacosta00} mention that the zeropoint for the
conversion from the F450W filter to Johnson B should be corrected by
$0.055$~mag. This correction was applied to the $B-V$ color of the
Andromeda photometry, leading to slightly redder colors.

The ground based observations were reduced using the
IRAF\footnote{IRAF is distributed by the National Optical Astronomy
Observatories, which are operated by the Association of Universities
for Research in Astronomy, Inc., under cooperative agreement with the
National Science Foundation.} standard CCD reductions and the {\tt
daophot} package \citep{stetson92} running under IRAF.

Details of the CTIO 4m BTC reductions of Sculptor can be found in
\citet{hurley99}. The data obtained with the Mosaic\,2 camera for
Sextans will be discussed in \citet{harbeck01}. Details of the
reduction of the CTIO 1.5m data of Carina will be described in
\citet{grebel01b}.

The color magnitude diagrams (CMD) for the LG dSphs are shown in
Fig.~\ref{fig_cmd}.

\section{Analysis \label{chap_analysis}}

To investigate the presence of a possible radial HB morphology
gradient for a given dSph we select the blue HB and red HB from the
CMD according to the selection boxes in Fig.~\ref{fig_cmd}. The
observations for the dSphs were made using different filter
combinations, therefore the color- and magnitude selection vary for
the galaxies. Even in the case of the five Andromeda dSphs, which were
all observed in the same filters, the HB morphology varies
significantly. Due to different contamination of the red HB by the RGB
we define the best selection criteria for each individual galaxy to
minimize contamination effects. A more quantitative analysis that
corrects for this contamination of the HB is performed in section
\ref{chap_hbindex}. The red edge of the BHB selection box and the blue
edge of the RHB box are chosen to avoid stars from the instability
strip. The boundaries of the instability strip in the different filter
sets are: $B-V=0.18$ to $0.4$ \citep{sandage90}; $V-I = 0.4$ to
$0.75$; $C-T_1=0.38$ to $0.64$ \citep{harbeck01}; $B-R=0.35$ to $0.75$
(based on visual inspection of Sculptor's CMD). In the case of Carina,
where a red clump population is visible, we select this region from
the CMD to compare its spatial distribution with that of the RHB.

Most of the dSphs in our sample have a non negligible ellipticity
$1-\case{b}{a}$ as listed in Tab. \ref{tab_obslog}. Considering just
the radial distance $\tilde{r}$ of a star to the center of its galaxy
would ignore this ellipticity and weaken any gradient effects. We
therefore calculate for each star an ellipticity corrected distance
$r$, equivalent to the major half axis. In the following we will
always use this corrected distance when talking about radial distances
to a galaxy center. The centers and position angles for the dSphs
where chosen from published work, as referenced in Table
\ref{tab_obslog}.  While different populations may exhibit different
centroids \citep{stetson97}, the offsets are small. We therefore
search for possible gradients in old and intermediate-age populations
by assuming azimuthal symmetry.

We use several different tools to analyze the radial distribution of
different populations:

\subsection{Cumulative radial distribution}

We plot the cumulative radial distribution of the BHB, RHB, and, if
present, the RC stars.  We then apply a Kolmogorov-Smirnov (K-S) test
to the radial cumulative distributions, which gives the probability of
whether the distributions of the BHB and RHB stars are drawn from the
same population.

\subsection{$T_2$ statistics of number ratios}
 We use the number ratio
\begin{eqnarray}
  N_{HB}=\frac{n_{BHB}}{n_{BHB}+n_{RHB}}
\end{eqnarray}
which measures the blueness of the HB \citep{mironov72}. $N_{HB}$
varies between 0 for a completely red HB and 1 for a totally blue
HB. $N_{HB}$ is calculated for radial bins with the width of a core
radius each. We overplot $N_{HB}$ for the individual bins in the
cumulative distribution diagrams as histogram bars
(Fig.~\ref{fig_rad}). If more than two core radii are covered by the
observations this method shows whether there is a trend in the
morphological gradient or just a local fluctuation.  We also calculate
$N_{HB}$ for all stars with a radial distance larger than a single
core radius. We refer to this radial selection as the ``outer
region'', while all stars within a core radius belong to the ``inner
region''. For Carina this analysis is also carried out to compare the
distribution of the RHB and the RC stars.

$T_2$ statistics, which quantify the probability that two number
ratios are derived from the same distribution \citep{dacosta96}, is
then performed on the two $N_{HB}$ values in the inner and outer
region. The $T_2$ value has a normal distribution around zero with a
width of one if the two compared number ratios are derived from the
same distribution. $T_2$ values therefore directly measure at what
$\sigma$-level two distributions are distinct. During the analysis it
turned out that the $T_2$-significance of an existing gradient is
strongly influenced by the selection of the radii within which
$N_{HB}$ is calculated. We therefore decided to separate the inner and
outer bin at the core radius for all dSphs to have a unique criterion,
even though one could easily tune up the $T_2$-significance by
individual bin selections for each dSph.

We also define the gradient strength parameter
\begin{eqnarray}
S[HB] = \frac{N_{HB}(r\le r_c)}{N_{HB}(r \ge r_c)}.  
\end{eqnarray}
$S[HB]$ compares the blueness of the HB in the inner and outer part
of a dSph, separated by the core radius, and is therefore a measure
for the strength of a morphology gradient on the scale length of a
core radius. $S[HB] =1$ for a homogeneous morphology without any
gradients.  $S[HB] < 1$ indicates a concentration of red HB stars,
while $S[HB] > 1$ for a concentration of the BHB stars. We define
$S[RC]$ in the same manner to compare the RC with the RHB.

\subsection{RGB radial distribution and metallicities \label{chap_rgbfe} }

We fit the mean ridge lines of the red giant branches (RGB) for all
the galaxies by a third order polynomial. This divides the RGB in a
red and a blue population (RRGB and BRGB). This ridge line is shifted
in the color direction to envelope the red and blue part of the RGB as
shown in Fig.~\ref{fig_cmd}. We tried to include as much of the RGB in
this selection as possible and to avoid field star contamination. The
same analysis as described for the HB stars is then applied to the RGB
stars.

\subsection{The HB  morphology index}

We determine the HB morphology index $$\frac{B-R}{B+V+R}$$
\citep{zinn86}. B, R, and V are the numbers of blue, red, and variable
stars in the HB, respectively. The determination of this index is
simple in the case of Sculptor, but becomes more complicated for
Sextans due to the field star contamination of the RHB. The HB of the
distant Andromeda companions and Tucana merges with the RGB, and the
number of RHB stars cannot measured directly. Since we need absolute
numbers of red HB stars to compare the HB morphologies of the dSphs we
need to correct for this contamination: The luminosity function (LF)
is plotted for all stars redder than the red end of the instability
strip (considered to be at B-V$=0.4$; V-I$=0.8$), as demonstrated in
Fig.~\ref{fig_hblum}. The HB is a clearly distinct peak above the
continuum of the RGB LF. We describe the LF of the RGB at the
luminosity of the HB by a linear function and subtract this from the
whole LF. Counting the remaining stars gives us a good estimate for
the correct number of red HB stars. The same method is used to correct
for the field star contamination in Sextans' HB, but with an
additional cut at the red end of the RHB. The HB morphology index is
determined for (i) the whole field covered, (ii) for radial bins with
the width of the core radius, and (iii) for the inner and outer bins.

\section{Radial gradients in the HB morphologies}

The radial cumulative distributions of the He core burning stars in
our dSphs are drawn in Fig.~\ref{fig_rad} with solid lines for the RHB
stars and with dashed lines for the BHB stars. The histogram bars show
the $N_{HB}$ values as a measurement of the HB morphology for radial
bins with the width of one core radius, while the thick bars show
$N_{HB}$ for all stars outside the core radius.

The corresponding statistical values (K-S-Test, $T_2$, and $S$) are
listed in Tab.~\ref{tab_stat}.  For Carina the results of the
comparison between the RC and the RHB are listed as well.

\subsection{The Milky Way companions} 

Due to the large angular extent of the nearby galaxies we have a
strong contamination with foreground stars, especially in Carina and
Sextans. \citet{ratnatunga85} estimate the expected field star
contamination for the globular cluster Pal\,3, which is close to
Sextans, to be $\sim 9.4$ stars in a $\pm 1$ magnitude bin per
arcmin$^2$ at the HB magnitude of Sextans. In contrast, for Sculptor
we expect only $1.3$ stars per arcmin$^2$. As can be seen in the CMDs,
this contamination affects only the RHB and RC number counts, which
will be overestimated. The foreground objects have a uniform
distribution and therefore let the contamined RHB and RC star sample
appear less concentrated than they actually are. The gradient
strengths determined here for the MW companions are therefore lower
limits.

All three Milky Way companions considered here have different HB
morphologies. Carina is the only one to show a prominent red
clump. All three show strong radial population gradients in the
morphology of the He-core burning star distribution.

The {\bf Carina} dSph shows no significant gradient in its HB
morphology: The K-S-probability that the RHB and BHB stars have the
same radial distribution is $~43$ percent; the $T_2$ value of $0.13$
is consistent with an equal distribution of RHB and BHB stars, as
well. But the RC stars show a very significant central concentration
compared to the HB. Such a concentration of the younger populations
was already suggested by \citet{mighell97}.

{\bf Sculptor} has the most significant HB morphology gradient in our
sample, as indicated by the high $T_2$-value of $6.2$ and a K-S-test
result of $1.8 \cdot 10^{-6}$. This result confirms the finding of
\citet{hurley99}, based on the same data, and that of \citet{majewski99},
who use independent observations. Care has to be taken when
interpreting the K-S test result due to the properties of the
observations with the BTC camera; this camera is a mosaic of 4 chips
with $5.4$ arcmin gaps. These gaps are smeared out by the elliptical
distribution of the stars. But one should keep in mind that the K-S
test is strictly defined for contiguous distributions only.

The HB morphology gradient in {\bf Sextans} is not as significant as
compared to Carina or Sculptor due to lower number statistics of the
sparse blue HB. But with $S=0.57$ and $T_2=2.6$ its gradient is still
very strong.

\subsection{The Andromeda companions}

The expected field star contamination towards Andromeda is $\sim 5.6$
stars per arcmin$^2$ in a $\pm1$ magnitude bin for the HB
luminosities. But due to the smaller apparent angular size of the
Andromeda dSphs, field star contamination does not play a major role
for the M\,31 dSph gradients.

The M\,31 dSphs show a variety of population gradients. {\bf Andromeda
I} seems to show a strong radial HB morphology gradient: While the
result of the K-S-test is consistent with no gradient, the result of
the $T_2=2.6$ and $S=0.37$ formally suggest a very strong
gradient. This is consistent with the result of
\citet{dacosta96}, whose work is based on the same data we used. But
the result for the outer bin is based on only 25 stars. This suggests
that the HB gradient may become more pronounced beyond the field
covered by HST and our result for Andromeda I is not reliable. A
larger field coverage would be desirable to clearly establish the
existence of a gradient, but no such data exist at present.

{\bf Andromeda II} does not show a significant gradient (K-S-test:
$22\%$, $T_2=1.2$).  This result is again consistent with the work of
\citet{dacosta00}. In {\bf Andromeda III} there is no obvious 
gradient (K-S-test: $33\%$, $T_2=0.9$) either.

The plot of the HB star distribution in {\bf Andromeda V} in
Fig.~\ref{fig_rad} suggests a central concentration of the red HB
stars. But the distributions of BHB and RHB stars are still consistent
with being similar (K-S-test: $24\%$, $T_2=0.7$).

Finally, in {\bf Andromeda VI} there is a significant HB gradient
(K-S-test: $0.1\%$, $T_2:2.7$).

\subsection{The lonesome Tucana}

Tucana has a very strong HB morphology gradient ($S=0.54$) at a high
significance level (K-S-test: $5\cdot 10^{-4}\%$, $T_2=4.4$).  While
all of the other nine dSphs are presumably bound to a massive spiral
galaxy, Tucana appears to be isolated within the LG. If Tucana has
always maintained a large distance from massive galaxies it would be
an example of undisturbed evolution in a dSph. If a HB gradient can
form in such an isolated environment, this may indicate that it is
created through internal processes in dSphs \citep{grebel00}.

\section{Metallicity gradients in dSphs seen in the RGB \label{chap_rgbgrad}}

We analyzed the radial distribution of the red and blue part of the
RGB as described in the Section \ref{chap_rgbfe}. For Carina the
contamination of the RGB by field stars is extremely high, and there
is no comparison field for statistical field star subtraction
available. It is therefore impossible to make a sensible analysis of
possible gradients: Small variations in the selection criterion for
the RRGB and BRGB can produce any desired gradient in the RGB
morphology. We therefore ignore the RGB of Carina in the further
analysis.

The resulting analysis of RGB gradients for the remaining sample of
dSphs is plotted in Fig.~\ref{fig_rgb} in the same way as in
Fig.~\ref{fig_rad} for the HB and listed in Tab.~\ref{tab_stat}.

There is a wide variety of RGB morphological gradients: Sculptor,
Sextans, Tucana, and Andromeda VI show strong and significant
concentrations of the RRGB stars. Andromeda II does not show any
change in its RGB morphology with increasing central distance.

In Andromeda I+III the blue RGB stars seem to be more
concentrated. For Andromeda I this effect has a low K-S-test
significance and can be attributed to small number statistics. Indeed,
if we reanalyze the HB and RGB gradient strength in the same way as
before, but separating at $60''$ instead of the core radius of $96''$,
the result is consistent with no gradients in both the HB and the
RGB. The apparent concentration of the blue RGB stars in Andromeda III
is also very low and the statistical significance of a possibly
existing gradient is also very low.

Because the dSphs, except for Carina, which is not considered here, do
not contain a significant intermediate age population, we interpret
red RGB stars as the more metal rich stars. A central concentration of
the red RGB stars would thus imply a metallicity gradient.

The gradient strengths for the HB ($S[HB]$) and for the RGB
($S[RGB]$) are calculated for each dSph with the same radial
binning. Therefore $S[HB]$ and $S[RGB]$ can be directly compared,
independent of the actual field coverage within a dSph. We compare
these two values in Fig.~\ref{fig_hbrgb}.  All dSphs that exhibit a
significant morphological gradient of the HB (Sextans, Tucana,
Sculptor, and Andromeda VI) always show a significant concentration of
the red RGB stars, too. There are no significant gradients in
Andromeda II, III, and V within the covered area. Note that the result
for Andromeda I (HB and RGB) may be unreliable. The nice correlation
between RGB and HB gradient strength visible in Fig.~\ref{fig_hbrgb}
strongly suggests that metallicity is an important driver of the
gradients.

\section{The dSphs in the metallicity-HB-index diagram \label{chap_hbindex}}

A useful tool to compare HB morphologies and metallicities is a plot
of metallicity vs. the HB index $\case{B-R}{B+V+R}$
\citep{zinn93,lee94,sarajedini97}. We calculate the HB index 
as described in the Section \ref{chap_analysis} for the eight
dSphs. The number counts for the blue, variable, and red HB stars are
listend in Tab.~\ref{tab_hbrgb} together with the resulting HB
index. We calculate the HB index for (a) the whole covered field and
(b) in bins with the width of one core radius. We found in the case of
Carina that we can not correct for the contamination of the RHB by RC
stars. It is also impossible to derive a mean metallicity for the old
HB forming population from the RGB only (what currently metallicities
are based on), since this metallicity would be dominated by
intermediate age stars. Putting Carina into the HB index diagram would
be not representative for its HB forming population. Carina is
therefore excluded from the investigation in this section.

\subsection{The whole covered field}

In Fig.~\ref{fig_leeplot} the mean HB-indices of the dSphs are plotted
versus the mean metallicities with open circles. The metallicities are
taken from the compilation of \cite{grebel00} except for Tucana and
some of the Andromeda companions. In Andromeda III, the spectroscopic
mean metallicity suffers from small number statistics \citep{guha01};
we adopt [Fe/H]$=-2.0$~dex \citep{arm93}. For Andromeda VI we derive
in Sect.~\ref{chap_leebin} [Fe/H] $= -1.6$~dex from our own analysis,
and [Fe/H] $\sim -1.6$~dex for Tucana, as well.  We also show the
locations of the MW globular clusters with data from
\citet{harris96} (filled circles) and for the four Fornax dSph
globular clusters based on data of \citet{buonanno99} (open
stars). Isochrones for three different relative ages are super-imposed
\citep{lee94}.

It turns out that the LG dSphs are not at the same location in this
diagram as the single age population globular clusters of the MW. In
particular the dSphs have redder HB morphologies than the mean of the
MW GC system, but there are a few very red HB, metal-poor GCs in the
MW that compare to the dSphs' HBs. If the HB morphology of the dSphs
is indeed determined by age and metallicity only, the location in this
diagram suggests that the dSphs may be systematically younger than the
MW globular cluster system. On the other hand, there is evidence from
deep main sequence photometry that the ages of nearby LG dSphs are
comparable to the oldest MW GC clusters, e.g. shown for Sculptor by
\citet{monk99}; for Carina by \citet{mighell97}. The [Fe/H]-HB-index 
diagram may not be the most reliable tool to compare relative ages of
the dSphs, since many parameters such as mass loss, [$\alpha$/Fe] and
helium content affect age determinations. Additionally, the star
formation histories of the dSph are by far more complex than that of
the GCs.

In particular the content of $\alpha$-elements in dSphs may be an
important parameter for the HB morphology. A recent spectroscopic
study by \citet{shetrone01} of RGB stars in the dSphs Draco, Ursa
Minor, and Sextans suggests a decreased $\alpha$-element abundance
($0.02 \le $[$\alpha$/Fe]$\le 0.12$) among these LG dSphs, compared to
the Galactic globular clusters. At the present time, the total effect
of $\alpha$-element abundances in the stellar envelope on the HB
morphology is not well understood. In general $\alpha$-elements,
especially oxygen, increase the opacity of stellar envelopes. For a
fixed mass, stars with increased [$\alpha$/Fe] will have lower surface
temperatures and therefore will form redder HB stars
\citep{VandenBerg00,VandenBerg01}. Accordingly, one would expect that
dSphs would form bluer HBs than the MW globular clusters. But an
enhanced opacity of the stellar envelope, e.g. due to an increased
[$\alpha$/Fe], should lead to an enhanced mass loss during the RGB
evolutionary phase. As a consequence, stellar populations with
increased [$\alpha$/Fe] might form lower mass HB stars with higher
surface temperatures. In this case the reduced $\alpha$-element
content of the dSphs could explain the systematically redder HBs of
the dSphs.

Nearly all globular clusters contain RGB stars that show abundance
patterns that indicate deep mixing, while the amount of deep mixing
stars can vary from cluster to cluster, as reviewed, e.g., in
\citet{kraft94}. Some models and correlation exist that suggest that
deep mixing may indeed be a second parameter in globular clusters
\citep{sweigart98,cavallo00}. If deep mixing were an internal second
parameter in dwarf spheroidals, one would expect different abundance
patterns among stars with different radial distances to the dSphs'
centers.

The five Fornax globular clusters are known to have very red HB
morphologies compared to their metallicity \citep{smith98,buonanno99}
and are located at similar positions as the dSphs in
Fig.~\ref{fig_leeplot}. According to \citet{buonanno99}, however, most
of the Fornax GCs have ages comparable to the MW GC system.

\subsection{Radial dependence of the HB index and the metallicity 
  \label{chap_leebin}}

Table \ref{tab_hbrgb} contains the HB morphology indices of the dSphs
for different radial bins. The HB morphological gradients detected
above are also reflected in the radial dependence of the HB morphology
index.

All dSphs observed with the HST have well observed RGBs in the
luminosity where star by star metallicity determinations using
fiducials are possible. In Sculptor the stars of the upper RGB are
saturated and we cannot perform a star by star analysis here. An
analysis of the mean color of the lower RGB was found to be too
unreliable to obtain meaningful results. In Sextans and in Carina
field star contamination would affect photometric mean
metallicities. Since in Carina there is a dominant intermediate age
population, the RGB color can not become translated into a
metallicity, since age becomes an important additional governing
parameter.

Tucana and Andromeda VI are the only two dSphs that have significant
HB morphology gradients in this sub-sample, where we can derive star
by star metallicities. To compare the location of different radial
bins in the [Fe/H]-HB-morphology diagram, we analyze the metallicity
distribution.  We use fiducials in the $V$,$I$ color space
\citep{dacosta90} and in the $B$,$V$ filters \citep{layden97} to
transform the location in the CMD of RGB stars into a metallicity. In
particular we use fiducials of the globular clusters M15, NGC 6752,
NGC 1851, and 47 Tuc with metallicities of [Fe/H] = $-2.17$~dex,
$-1.54$~dex, and $-1.29$~dex, and $-0.71$~dex according to the sample
of \citet{dacosta90}.

The upper panels of Fig.~\ref{fig_fedist} show the metallicity
distributions for the inner and outer parts as well as for the whole
covered field of Andromeda VI and Tucana. In the lower panels of the
same figure the error-folded distribution of metallicities is
shown. We assumed a general uncertainty of $\pm0.05$~dex in [Fe/H] plus
an additional photometric error folded with the dependence
$\case{\Delta [\mathrm{Fe/H}]}{\Delta (B-V)}$ of the metallicity on
the a star's color.

From the [Fe/H] distributions of Andromeda VI and Tucana we calculate
the median as an estimate for the mean metallicities. Although the
median is quite robust against outliers, we exclude stars with
metallicities higher than $-1.0$~dex, which we consider to be field
stars. Another source of contamination for [Fe/H]$\ge-1$~dex could be
CH stars. Both for Tucana and Andromeda VI we derive a mean [Fe/H]
$=-1.6$~dex. Note the bimodal metallicity distribution in Tucana. The
derived [Fe/H] for Andromeda VI is compatible with the result of
\citet{arm99}, but $0.3$~dex more metal-poor than the result of
\citet{grebel99}. Our analysis estimates Tucana to be $0.2$~dex more
metal-rich than the study by
\citet{saviane97}.

The visual inspection of the metallicity distributions of Andromeda VI
and Tucana (Fig.~\ref{fig_fedist}) reveals that the ratio of metal
poor and metal rich stars changes significantly for different radial
selections. For Andromeda VI we derive metallicities of $-1.55$~dex
and $-1.7$~dex for the inner and outer region, respectively. From
Table \ref{tab_hbrgb} HB indices of $-0.85$ and $-0.55$ are obtained
for the same regions. The metallicities of the inner and outer bins of
Tucana are calculated to be $-1.56$~dex and $-1.62$~dex. The HB
indices of the bins are $-0.27$ and $0.01$, respectively. It is worth
mentioning that the strong RGB morphology gradients that we derived in
section \ref{chap_rgbgrad} for Tucana and Andromeda VI can be
accounted for by a relative small difference in the mean metallicity.

\subsubsection{Reality check}

We will briefly discuss three points regarding the reality of the
apparent gradient in the metallicity in Andromeda VI and Tucana and
the influence of field star contamination:

\begin{enumerate}

\item  Keck/LRIS spectroscopy of bright giants in four Andromeda dSphs
   \citep{guha01} directly gives the contamination fraction from a
   radial velocity membership criterion.  The only non-members found
   (M31 field giants and foreground Galactic dwarf stars) are
   invariably found well away from the center of the dSphs. In fact,
   the cross-identified stars in common between the HST star list and
   the Keck spectroscopy list does not contain even a single
   non-member - i.e. zero non-members out of a total of 34 cross-ID-ed
   stars in the four dSphs.  The HST star-by-star metallicity
   estimates (for which we are considering contamination issues) are
   dominated in number by RGB stars fainter than those in the Keck
   spectroscopic sample, but this only means that the fractional
   contamination from foreground Galactic dwarfs will be even lower.

\item  Contamination, both by foreground Galactic dwarf stars and M31
   field RGB stars (halo or disk) tends to cause an apparent increase
   in the estimated [Fe/H] of the population.  Since contamination is
   bound to be greater in the outer ($r \ge r_c$) sample than in the
   inner sample, this goes the wrong way to explain the [Fe/H] radial
   gradient that is observed.

\end{enumerate}

\subsubsection{The inner and outer bin in the HB-index-[Fe/H] diagram}

The location of the inner and outer bins of the two dSphs in the
[Fe/H]-HB-index diagram is plotted in Fig.~\ref{fig_leeplot}. The
positions of the inner and outer bins of the two galaxies in this
diagram suggest that metallicity is the main parameter driving the HB
gradient. According to the model of \citet{lee94} there is no {\em
internal} second parameter required to explain the HB morphology
gradients in these two galaxies. But taking into account the
uncertainty of the HB morphology index (especially to mention the
contamination of the HB by the RGB and its correction and the lack of
direct RR Lyrae identification), an age spread of 2~Gyr in these two
galaxies can not be excluded.

\subsubsection{The bimodal metallicity distribution of Tucana}

The plot of Tucana's metallicity distribution in Fig.~\ref{fig_fedist}
suggests a bimodal distribution; in fact that the multimodality exists
in the inner and the outer bin, with similar peaks, supports the
reality of this effect.

This multimodal metallicity distribution suggests that Tucana formed
stars in two or more major events. In the previous section we
demonstrated that there is no evidence for a significant age gradient
in Tucana. Furthermore, the absence of carbon stars in Tucana
\citep{battinelli00} as well as the absence of a red clump 
indicates that there is no significant intermediate age population.
The star formation in Tucana therefore happend in a quite narrow age
window at least $10$~Gyr ago. Different radial distributions of the
populations that formed in the different epochs are still visible at
the present time by the morphological gradients in the RGB and the HB.

\section{Global parameters governing the morphology gradients in the HB \label{param}}

To investigate the possible impact of environmental effects, we plot
the gradient strength $S[HB]$ of the dSphs against their absolute
magnitude $M_V$ and against their deprojected distance to the nearest
massive spiral galaxy (Fig.~\ref{fig_prop} a and b). The absolute
magnitude and distances were taken from the compilation of
\citet{grebel00}. As pointed out by \citet{bellazzini96}, one should
keep in mind that the currently observed distance of the galaxy may
not be representative of its orbit and the distance during the star
forming episodes; the probability to find a satellite galaxy in its
apogalactic orbit position is higher than at the pericenter.  The
whole sample of galaxies is quite inhomogeneous in gradient strengths
and does not show a significant dependence of $S[HB]$ on the radial
distance or the absolute luminosity.

The field coverage of our data compared to the radial extent of the
dSphs in our sample varies a lot. But for each of our sample galaxies
the data cover an area of at least $1.25 \cdot r_c$. We therefore
redetermined $S[HB]$ for all galaxies taking only the central $1.25$
core radii into account. This gradient strength will be called
$S_c[HB]$. These new gradient strengths $S_c[HB]$ and their
significance $T_2$ are listed in Tab.~\ref{tab_core}.  Within this
common central radius of $1.25\cdot r_c$ the gradient strengths for
the Milky Way companions and Tucana tend to become weaker (reflected
by slightly increased $S_c[HB]$) than before and closely resemble the
strength of the M\,31 dSphs measured within the same core radius. The
case of Andromeda V demonstrates that an insufficient field coverage
may lead to completely different results. We expect that the HB
morphology variations in the M\,31 companions may become more
pronounced and significant if a larger area of these galaxies were
covered.

A new comparison of the $S_c[HB]$ parameters with the luminosity and
distance to the nearest spiral does not reveal any significant
correlations either. We therefore exclude the absolute luminosity and
the present day radial distance from the nearest massive spiral as
parameters governing the HB gradients within the restrictions of our
sample.

\section{Conclusions}

We have compared the HBs of nine Local Group dSphs. The existence of a
morphological gradient of the HB turns out to be a common, but not a
defining, feature of dSphs. If there is a population gradient, the RHB
or RC stars are always more concentrated than the BHB stars; there is
no counter example in our sample. The main parameters governing the HB
morphology are age and metallicity. There are other possible
candidates such as stellar rotation, but there is no obvious mechanism
that would systematically alter this parameter in the low-density
environment of a dSph as a function of radius.

Increasing the helium abundance would lead to lower mass HB stars and
would therefore foster the formation of bluer HBs \citep{lee94}. For
the solar vicinity a correlation between metallicity and helium was
found with $\Delta$Y$/\Delta$Z$\sim3$ \citep{pagel98}. Increased
[Fe/H] should therefore be accompanied by an enhancement of [Y/H] and
might overall reduce the effect of metallicity on the HB.

Metallicity as one defining parameter for the HB morphology could
produce morphology gradients through a much more efficient
self-enrichment of the star forming gas in the central region of a
dSph.  The driver of the gradient could be the gravitational potential
of the dSph which would either retain the star forming gas longer in
its central regions, or retain both the gas and newly formed metals
\citep{mayer01}. The correlation between the central concentration of red RGB
stars (supposed to be metal-rich) and the concentration of red HB
stars (also assumed to be metal-rich) in Fig.~\ref{fig_hbrgb},
supports the idea of a global metallicity gradient.  Examples where
metallicity is an important driver for the HB morphology gradients in
our sample are Sextans, Sculptor, Tucana, and Andromeda VI. The
location of the inner and outer bins of Andromeda VI and Tucana is
consistent with a pure metallicity effect.

In contrast, age as a governing parameter of the HB morphology would
support the idea that gas in the dSphs could be retained in the center
of the dSphs for a significantly longer time than in the outskirts
\citep{grebel00}. In our sample there is no evidence that age is 
the dominant governing parameter for the HB gradients. At least in the
Carina dSph age is a viable parameter that correlates with the
gradient of the He-core burning stars, as indicated by the strongly
concentrated intermediate age RC. It is very interesting that the HB
in Carina does not show a gradient effect. But Carina is the only dSph
in our sample that has a dominant intermediate-age population and its
star formation history is not directly comparable to the remaining
predominantly old dSphs.

The central concentration of the youngest stars has also been seen in
Fornax and in the transition type galaxy Phoenix
\citep{stetson98,martinez99}. One may consider the transition type
galaxy Phoenix as an example of a progenitor of a dSph with a strong
population gradient.

Overall, in this study we were able to find dSphs with indications of
(a) metallicity gradients (Tucana, Andromeda VI, Sextans, and Sculptor)
(b) no or weak gradients (Andromeda I, II, III, and V) and (c) age
gradients (Carina). Our observations seem to confirm the picture that
no two galaxies are alike, not even when they are of the same
morphological type.

We found no significant evidence that the proximity to a massive
galaxy influenced the formation of an HB gradient, in contrast to van
den Bergh's (1994) and Grebel's (1997) correlation between a dwarf
galaxy's stellar content and its distance to the Milky Way or
M\,31. The HB and RGB morphology gradient in Tucana is an interesting
result due to the fact that this galaxy is not associated with any
massive spiral galaxy. It also exhibits a multimodal metallicity
distribution, indicative of multiple episodes of star formation. If
this dSph was not close to a massive galaxy while it formed stars,
environmental effects like tidal gas stripping or an external UV field
from a nearby galaxy rather than an extragalactic UV field, could not
have introduced the population gradient as far as the progenitors of
the HB stars are concerned. Hence it looks as if the formation of
population gradients in dSphs is a process that is caused by interior
processes within a dSph or that is triggered by external events that
occur independently of a nearby massive galaxy.

What could trigger or inhibit the formation of an HB gradient in
dSphs? If the formation of a population gradient is the normal mode of
dSph evolution it may, e.g., have been stopped due to collisions with
intergalactic gas clouds in some glaxies. Such a collision may strip
the remaining gas out of the evolving galaxy and end any further star
formation. On the other hand an intergalactic cloud may have been
captured by a dSph and caused a second star forming epoch, a scenario
that is suggestive for Tucana. Yet the shallow potential of a dSph
makes it difficult to keep intergalactic gas clouds bound
\citep{hirashita99}. \citet{maclow99} found that the amount of gas
lost by dwarf galaxies during their star forming epoch depends on the
luminosity of the star burst and on the parent galaxy mass. The
strength of the star forming time may therefore have been influenced
by the starburst itself.

There is also no evidence that the spatial scales on which the
population gradients appear are correlated with the core radius of a
dSph. The minimum scale on which a population gradient becomes evident
can vary a lot. Gradients seem to become most evident at a field
coverage of at least one or two core radii. On smaller scales
gradients may disappear (e.g., Andromeda I). But the HB morphology
gradient in Sculptor can be seen down to a radius of $4$''
($2/3\,r_c$) with a K-S-Test probability of $5$~\%. To obtain a
measure of the scale length for existing morphology gradients requires
a field coverage of order of the tidal radius, which is lacking for
many galaxies at present. With a sufficient field coverage one could
define core radii for the RHB and BHB populations and attempt to
correlate them with a natural length scale of a dSph.

\acknowledgments

This work is partly based on observations made with the NASA/ESA
Hubble Space Telescope, obtained from the data archive at the Space
Telescope Science Institute. STScI is operated by the Association of
Universities for Research in Astronomy, Inc. under NASA contract NAS
5-26555.

The Digitized Sky Surveys were produced at the Space Telescope Science
Institute under U.S. Government grant NAG W-2166. The images of these
surveys are based on photographic data obtained using the Oschin
Schmidt Telescope on Palomar Mountain and the UK Schmidt
Telescope. The plates were processed into the present compressed
digital form with the permission of these institutions. The Second
Palomar Observatory Sky Survey (POSS-II) was made by the California
Institute of Technology with funds from the National Science
Foundation, the National Geographic Society, the Sloan Foundation, the
Samuel Oschin Foundation, and the Eastman Kodak Corporation.  The
Oschin Schmidt Telescope is operated by the California Institute of
Technology and Palomar Observatory.  The UK Schmidt Telescope was
operated by the Royal Observatory Edinburgh, with funding from the UK
Science and Engineering Research Council (later the UK Particle
Physics and Astronomy Research Council), until 1988 June, and
thereafter by the Anglo-Australian Observatory.

D.G. acknowledges financial support for this project received from
CONICYT through Fondecyt grant 8000002, and from the Universidad de
Concepci\'on through research grant No. 99.011.025. J.H. is a Cottrell
Scholar of Research Corporation and acknowledges their support. We
would like to thank A. Burkert and J. S. Gallagher for helpful
discussions.

\begin{figure}
\figcaption{ Demonstration 
      of the influence of age and metallicity on the RGB morphology:
      We show isochrones \citep{girardi00} for $z=0.004$ and
      $z=0.0004$ for ages of $6$, $10$, and $14$~Gyr. \label{fig_iso}}
\end{figure}

\newpage

\begin{figure*}
%
\figcaption{ Digital
   Sky Survey images of the nine dSphs in our sample with the
   footprint of the covered field. \label {fig_chart}}
\end{figure*}

\newpage

\begin{figure*}
%
\figcaption{Color
 magnitude diagrams of all dwarf spheroidals in our sample. The blue
 and red horizontal branch stars,  and the red clump stars in Carina are
 selected according to the  boxes. \label{fig_cmd}}
\end{figure*}

\newpage

\begin{figure}
%
\figcaption{As an
   example for counting HB stars in the HST photometry where the red
   HB and the RGB merge, the luminosity function (LF) for all stars
   with $B-V \ge 0.4$ is shown. At the luminosity of the HB the LF is
   modelled by a linear function und subtracted from the total. The
   remaining HB stars form a clearly distinct peak above the zero
   count level. \label{fig_hblum}}
\end{figure}

\begin{figure*}
%
\figcaption{ Radial 
  distributions of the blue (dotted line) and red (solid line)
  horizontal branch (HB) stars for all our sample dSphs.  For Carina
  we plot in addition the distribution of the red clump stars with a
  dashed line. The selection of the population is made according to
  Fig.~\ref{fig_cmd}. For all of the Milky Way satellites a
  concentration of the red HB or the red clump stars (Carina) can be
  seen. The histogram bars represent the blueness index $N_{HB}$ for
  the radii covered by the bars. The vertical line marks the core
  radius. The strong horizontal line outside the core radius indicates
  the mean HB ratio in the outer region, if more than two core radii
  are covered.  \label{fig_rad}}
\end{figure*}
\clearpage

\begin{figure*}
%
\figcaption{ Radial
  distributions of the blue (dotted line) and red (solid line) part of
  the red giant branch (RGB) stars for all our sample dSphs.  Due to
  the strong field star contamination we can not analyze the radial
  distribution of the RGB stars in the Carina dSph.  The selection of
  the red and blue RGB populations is made according to
  Fig.~\ref{fig_cmd}. The histogram blocks and numbers have the same
  meaning as in Fig.~\ref{fig_rad}. \label{fig_rgb}}
\end{figure*}
\newpage

\begin{figure}
%
\figcaption{Comparison of the gradient strengths $S$ for 
  the horizontal branch stars ($S[HB]$) and for the red giant branch
  stars ($S[RGB]$).  Galaxies with $S[RGB] \le 1$ show a central
  concentration of red RGB stars, while galaxies with $S[RGB]\ge1$
  have a central concentration of blue RGB stars. The gradient
  strengths for those galaxies with $S[RGB]$ slightly larger than $1$
  is still consistent with no gradient effects.  There is a clear
  correlation between the gradients in these two populations:
  According to Tab.~\ref{tab_hbrgb}, Sex, Tuc, And VI and Scl have
  significant gradients both in the RGB and HB. And V has weak
  gradients, while And II and And III don't show significant
  gradients. And I suffers from low number statistics.
  \label{fig_hbrgb}}
\end{figure}

\clearpage

\begin{figure*}
%
\figcaption{ a) Dependence 
 of the population gradient as indicated by the HB gradient index
 $S[HB]$ (see text for definition) on the absolute luminosity of the
 dSph. $S[HB]=1$ corresponds to an equal distribution of the blue and
 red HB stars, while $S[HB]=0$ would indicate the maximal
 concentration of the red HB stars. b) $S[HB]$ compared to the
 logarithm of the distance to the nearest massive spiral galaxy. Milky
 Way companions are plotted with open circles, Andromeda satellites
 with filled dots. The distant galaxy Tucana is plotted with the open
 rectangle. There is no obvious correspondence between these
 parameters. \label{fig_prop} \label{fig_core}}

\end{figure*}

\newpage
\begin{figure}
%
\figcaption { The Local Group dSphs in the HB 
  index -- metallicity plot (open circles).  The dSphs' HBs appear
  systematicaly redder for a given metallicity compared to the Milky
  Way globular clusters (filled circles). This suggests the dSphs may
  be slightly younger than the GCs.  The location of the Fornax
  globular clusters (open stars) in this diagram indicates they are
  more comparable to the LG dSphs than the MW globulars. For the two
  dSphs Andromeda VI and Tucana the mean metallicity and HB indices
  for the inner and outer regions are drawn with small circles and
  connected with a solid line. For these two galaxies the HB
  morphology gradient can be explained by pure metallicity
  effect.\label{fig_leeplot}}
\end{figure}

\newpage
\begin{figure*}
%
\figcaption{Metallicity distribution
  of RGB stars in Andromeda VI and Tucana. We show the histogram of
  the [Fe/H] distribution in the upper half: The dotted bars show the
  distribution of all stars, while the solid line counts stars within
  a core radius and the dashed line counts stars more distant than a
  single core radius. We derive for the inner and outer regions
  metallicities of $-1.55$~dex and $-1.7$~dex for Andromeda VI;
  $-1.56$~dex and $-1.62$~dex for Tucana. In the lower part of the
  diagrams show the normalized density distribution of the [Fe/H]
  measurements folded by their errors for the same radial selections
  as for the upper panels. Note that the amplitude of these
  distributions does not reflect the number of stars. The radial
  [Fe/H] gradients still come out very nicely. Even the bimodal [Fe/H]
  distribution in Tucana is still present. \label{fig_fedist} }
\end{figure*}

\begin{deluxetable}{llcccccl}
\tablecaption{Observing log \label{tab_obslog}}
\tabletypesize{\small}
\tablewidth{0pt}
\tablehead{
\colhead{Galaxy} &
\colhead{Telescope, Instrument} &
\colhead{Filters} &
\colhead{$r_o$['] \tablenotemark{a}} &
\colhead{$r_c$ \tablenotemark{a}} &
\colhead{$r_t$ \tablenotemark{a}} &
\colhead{$1-{a\over b}$} &
\colhead{Reference}
}
\startdata
Carina   & CTIO 1.5m          & C,R      & $25$'  & $8.8$' \tablenotemark{b} & $29$'  & $0.33$  & \citet{grebel01b}\\
Sculptor & CTIO 4m, BTC       & B,R      & $30$'  & $5.8$' \tablenotemark{b} & $76$'  &  $0.32$ & Hurley-Keller et al. (1998)\\
Sextans  & CTIO 4m, Mosaic\,2 & C, T$_1$ & $40$'  & $16.6$'  \tablenotemark{b} & $160$'& $0.35$ & \citet{harbeck01}\\
Tucana   & HST, WFPC\,2 & F555W, F814W   & $1.7$' & $0.7$' \tablenotemark{c} & \nodata     & $0.48$ & (HST archive) \\
And I    & HST, WFPC\,2 & F439W, F555W   & $2$'   & $1.6$' \tablenotemark{d} & $13.3$'  & $0$    & \citet{dacosta96}\\
And II   & HST, WFPC\,2 & F439W, F555W   & $2.8$' & $1.6$' \tablenotemark{d} & $17$'    & $0.3$  & \citet{dacosta00}\\
And III  & HST, WFPC\,2 & F439W, F555W   & $2$'   & $1.3$'  \tablenotemark{d}& $6.2$'   & $0.6$  & (HST archive)\\
And V    & HST, WFPC\,2 & F439W, F555W   & $2$'   & $0.5$' \tablenotemark{e} & \nodata  & $0$    & (HST archive)\\
And VI   & HST, WFPC\,2 & F439W, F555W   & $2$'   & $1.3$' \tablenotemark{e} & \nodata  & $0.23$ & (HST archive)
\enddata
\tablecomments{ \tablenotemark{a}~ $r_o$: radius covered by observations,
  $r_c$: core radius, $r_t$: tidal radius. 
\tablenotemark{b}~ \citet{irwin95} \tablenotemark{c}~ \citet{saviane96}
 \tablenotemark{d}~ \citet{caldwell92}
\tablenotemark{e}~ \citet{caldwell99}}

\end{deluxetable}

\newpage

\begin{deluxetable}{lccccccccc}
\tablecolumns{10}
\tablewidth{0pt}

\tablecaption{Statistical significance of population gradients
    \label{tab_stat}}

\tablehead{
\colhead{} & 
\multicolumn{3}{c}{horizontal branch} & 
\multicolumn{3}{c}{red clump} &
\multicolumn{3}{c}{red giant branch}\\

\colhead{Galaxy} &
\colhead{K-S [\%]} &
\colhead{$T_2$} & 
\colhead{$S[HB]$} &
\colhead{K-S [\%]} &
\colhead{$T_2$} &
\colhead{$S[RC]$} &
\colhead{K-S [\%]} &
\colhead{$T_2$} &
\colhead{$S[RGB]$}\\
}
\startdata
Carina   & $43$  & $0.13$    & $1.0$ & $5\cdot 10^{-4}$ & $ 3$ & $0.57$ & \nodata & \nodata & \nodata \\
Sculptor & $1.8\cdot 10^{-9}$ & $6.2$ & $0.65$  & \nodata & \nodata  & \nodata & $5\cdot 10^{-3}$ & $4.5$ & $0.86$\\
Sextans  & $0.5$  & $2.6$   & $0.57$ & \nodata & \nodata & \nodata  & $1.5 \cdot 10^{-4}$ & $5.3$ & $0.81$ \\
Tucana   & $5\cdot 10^{-4}$ & $4.4$  & $0.54$ & \nodata & \nodata & \nodata  & $5$ & $2.9$ & $0.72$   \\
And I    & $44$ & $2.6$    & $0.37$ & \nodata & \nodata & \nodata   & $49$ & $2.9$ & $1.3$   \\
And II   & $22$  & $1.2$    & $0.72$ & \nodata & \nodata & \nodata  & $99$ & $0.9$ & $1.1$ \\
And III  & $33$  & $0.9$    & $0.71$ & \nodata & \nodata & \nodata  & $24$ & $0.5$ & $1.05$ \\
And V    & $24$  & $0.7$    & $0.83$ & \nodata & \nodata & \nodata  & $6$  & $1.3$ & $0.9$ \\
And VI   & $0.1$  & $2.7$    & $0.56$ & \nodata & \nodata & \nodata & $8 \cdot 10^{-4}$  & $6.6$  & $0.74$ \\
\enddata
\end{deluxetable}

\newpage

\begin{deluxetable}{lcccc}
\tablecolumns{5}
\tablewidth{0pt}
\tabletypesize{\small}
\tablecaption{HB morphology indices
    \label{tab_hbrgb}}

\tablehead{
\colhead{Galaxy} &
\colhead{\#B} & 
\colhead{\#V} &
\colhead{\#R} &
\colhead{$\case{B-R}{B+V+R}$}
}
\startdata
Carina   & \nodata & \nodata & \nodata & \nodata \\ \hline \hline
Sculptor \tablenotemark{a} &  353 & 243 & 299 & 0.06 \\ \hline
  I\tablenotemark{b}	& 136 & 98 &  184  &-0.11\\
  II	& 139 & 89 &   98  & 0.13\\
III	& 19& 7 & 8 & 0.32\\
IV	&  27& 18& 6 & 0.41\\
V	& 20& 17& 2 &  0.46\\
VI 	& 5  & 10 & 0 & 0.33\\
VII 	& 4  &  1 & 0  & 0.8 \\ \hline \hline

Sextans  & 102 & 24 & 237 & -0.37 \\ \hline 
 I	& 29 & 6 & 120 & -0.59 \\
 II	& 58 & 13 & 104 & -0.26 \\
 III	& 12 & 3 & 25 & -0.32 \\
 IV	& 3 & 2 & 4 & -0.11\\ \hline \hline
Tucana   & 162 & 58 & 264 & -0.2 \\ \hline
 I	& 48 & 23 & 133 & -0.42 \\
 II	& 65 & 25 & 79 & -0.08 \\
 III	& 37 & 7 & 36 & 0.01 \\
 IV	& 9 & 2 & 9 & 0 \\ \hline \hline
And I    &  91 & 88 & 1144 & -0.8 \\ \hline
 I	& 88 & 85 & 1079 & -0.8 \\
 II	& 3 & 3 & 31 & -0.75 \\ \hline \hline 
And II   & 140 & 55 & 923 & -0.7 \\ \hline
 I	& 90 & 36 & 614 & -0.7  \\ 
 II 	& 50 & 19 & 247 & -0.62  \\ \hline \hline
And III  & 78 & 81 & 552 & -0.67 \\ \hline 
 I 	& 39 & 44 & 328 & -0.7 \\
 II	& 34 & 31 & 196 & -0.61 \\
 III	& 5 & 6 & 51 & -0.74 \\ \hline \hline
And V    & 89 & 111 & 562 & -0.62 \\ \hline
  I	& 19 & 25 & 169 & -0.7 \\
  II	& 40 & 41 & 244 & -0.63 \\
  III	& 29 & 42 & 153 & -0.55\\
  IV	& 3 & 3 & 8 & -0.44\\ \hline \hline
And VI   & 177 & 308 & 1749 & -0.7 \\ \hline
 I 	& 114 & 208 & 1445 & -0.85 \\
  II	& 63 & 100 & 315 & -0.53	
\enddata
\tablecomments{ \tablenotemark{a}~ 
  The first line of a dSph contains number counts for the whole
  covered field.  \tablenotemark{b}~ The numbers I, II, \ldots refer
  to number counts in the first, second, etc. radial bins.}

\end{deluxetable}

\newpage

\begin{deluxetable}{lcc}
\tablecolumns{3}
\tablewidth{0pt}

\tablecaption{HB gradients in the inner $1.25\cdot r_c$
    \label{tab_core}}

\tablehead{
\colhead{Galaxy} &
\colhead{$T_2$} & 
\colhead{$S_c[HB]$}
}
\startdata
Carina   & $1.4$ & $1.8$ \\
Sculptor & $2.1$ & $0.81$ \\
Sextans  & $2.4$ & $0.54$ \\
Tucana   & $2.0$ & $0.66$ \\
And I    & $2.6$ & $0.37$ \\
And II   & $1.6$ & $0.65$ \\
And III  & $0.7$ & $0.73$ \\
And V    & $0.3$ & $1.2$ \\
And VI   & $2.7$ & $0.56$ \\
\enddata

\end{deluxetable}

\end{document}